\documentstyle[11pt,newpasp,twoside,epsfig]{article}
\def\edcomment#1{\iffalse\marginpar{\raggedright\sl#1\/}\else\relax\fi}
\reversemarginpar

\begin{document}
\title{Implications of modes of star formation for the overall dynamics of
galactic disks}
 \author{Burkhard Fuchs}
\affil{Astronomisches Rechen--Institut, M\"onchhofstr.~12-14, 69120 Heidelberg,
Germany}

\begin{abstract}
One of the present concepts for the onset of massive star formation is the
Kennicutt criterion. This relates the onset of massive star formation to a
general gravitational instability of the gas disks of spiral galaxies. It is 
often overlooked, however that such gravitational instabilities of the gas 
disks have severe implications for the overall stability of the gas {\em and} 
star disks of spiral galaxies. I show by numerical simulations of the evolution 
of a combined gas and star disk that the violation of the stability condition 
induces violent dynamical evolution of the combined system. In particular the 
star disk heats up on time scales less than a Gyr to unrealistic high values of
the Toomre stability parameter $Q$. The morphologies of both the star and gas 
disk resemble then no longer observed morphologies of spiral galaxies. Star 
formation of stars on low velocity dispersion orbits would lead to dynamical
cooling of the disks to more realistic states. However, the required star 
formation rate is extremely high.
\end{abstract}

\section{Massive star formation and disk stability}

In an influential study Kennicutt (1989) has introduced the
concept that massive star formation in spiral galaxies is related
to a general gravitational instability of the gas disks of
the galaxies. He demonstrated for a sample of Sc galaxies that the
inner parts of the disks with massive star formation, which is
observed in the form of numerous HII regions, have sharp outer
radial boundaries, which coincide with the threshold of
gravitational instability of the gas disks. Doubts about this
concept were raised by Ferguson et al. (1996, 1998). They
have shown that this transition is not as sharp as claimed by
Kennicutt (1989), because they detected HII regions also in the
outer parts of the disks of some of the galaxies in Kennicutt's
sample. Furthermore they have argued that the gas disk of the
prototype galaxy of the sample, NGC\,6946, might not have reached the
threshold of instability at all.

It is the aim of the present paper to point out implications of the concept 
of gravitational instability of the gaseous components of galactic disks for 
the overall stability of galactic disks.

Since the pioneering works of Safronov (1963) and Toomre (1964) there is a rich
literature on disk stability which cannot be reviewed here in its entirety, but
I recall only the basic principles. The gravitational stability of rotating,
self gravitating disks is regulated by two effects. First there is the well known
Jeans instability, which implies that perturbations of the disks with scales
larger than the Jeans length are dynamically unstable,
\begin{equation}
\lambda > \lambda_{\rm J} = \frac{\sigma^2}{G \Sigma}\,,
\end{equation}
where $\sigma$ denotes the turbulent velocity dispersion in the case of a gas
disk, or the stellar velocity dispersion in case of a star disk. $\Sigma$
denotes  the surface density of the disk and $G$ is the constant of gravity. In
a rotating disk, however, there is an upper limit to the scales on which
perturbations can grow. When a patch of a disk shrinks, its detailed angular
momentum referred to the center of the patch is conserved. This leads to
additional centrifugal forces tearing the patch apart, and perturbations on 
scales larger than
\begin{equation}
\lambda > \lambda_{\rm c} = \frac{G \Sigma}{\Omega^2}\,,
\end{equation}
where $\Omega$ denotes the angular velocity of the disk, are suppressed. Obviously, if the
upper limit is smaller than the lower limit,
\begin{equation}
\lambda_{\rm c} < \lambda_{\rm J}\,,
\end{equation}
the disk is stable on all scales. If the rhs of equation (3) is divided by the
lhs, the stability criterion can be expressed by a single number, the famous
Toomre $Q$ stability parameter. It is derived by searching for ring--like
solutions of either the hydrodynamical, the Jeans equations or the Boltzmann
equation in the case of a star disk combined with the Poisson equation in order
to take account of the self--gravity of the perturbations. It can be shown
that the stable solutions are separated from the unstable, growing solutions in
a parameter space spanned by the radial wave lengths of the ring--like density
perturbations expressed in terms of the critical wave length, $\lambda_{\rm
crit}=4 \pi^2 G \Sigma / \kappa^2$, and the stability parameter $Q$,
\begin{equation}
Q = \frac{\sigma_{\rm U}\kappa}{\alpha G \Sigma}\,,
\end{equation}
where $\sigma_{\rm U}$ denotes the turbulent velocity dispersion of the gas 
or the radial stellar velocity
dispersion and $\kappa$ is the epicyclic frequency, $\kappa = \sqrt{2} \Omega
\sqrt{2+\frac{r}{\Omega}\frac{d\Omega}{dr}}$. $\alpha$ is a numerical
coefficient, which is equal to $\pi$ for an isothermal gas disk and ranges 
between 3.6 and 3.9 for a star disk depending on the exact form of the velocity
distribution (Toomre 1964, Fuchs \& von Linden 1998). This is illustrated in
Fig.~1 for the simplest case of an isothermal gas disk. The line separating the
stable from the unstable solutions is given by (cf.~Binney \& Tremaine 1987)
\begin{equation}
1 = \frac{\lambda_{\rm crit}}{|\lambda |}\frac{1}{1+\frac{Q^2}{4}\left( 
\frac{\lambda_{\rm crit}}{\lambda}\right)^2}\,.
\end{equation}
\begin{figure} [h]
\begin{center}
\epsfxsize=5cm
   \leavevmode
     \epsffile{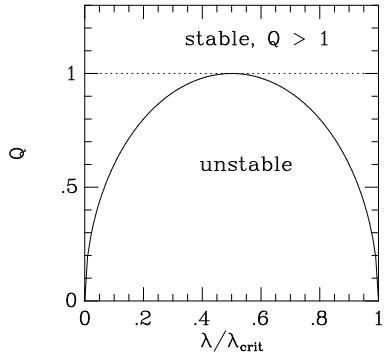}
\caption{Separation of unstable from stable ring--like density perturbations of
an isothermal gas disk in the parameter space spanned by the radial wave lengths
$\lambda$ of the density perturbations and the dimensionless $Q$ stability
parameter. The wave lengths are expressed in terms of the critical wave length
$\lambda_{\rm crit}$}
\label{fig1}
\end{center}
\end{figure}
As can be seen from from Fig.~1 the disk is stable on all scales, if 
\begin{equation}
Q \geq 1\,.
\end{equation}
The criterion for stability becomes more complicated, if the disk consists of
more than one component (Biermann 1975, Jog \& Solomom 1984, Romeo 1992,
Elmegreen 1995, Fuchs \& von Linden 1998). I shall not go here through a
mathematical
derivation, but illustrate the main effect by a `Gedanken'--experiment. Suppose
one splits a disk into two equal halves, which means according to their
definitions that $\Sigma_{1,2} = \frac{1}{2} \Sigma$, $\lambda_{\rm crit\,1,2}
 = \frac{1}{2}\lambda_{\rm crit}$, and $ Q_{1,2} = 2 Q$. The surface separating stable from
 unstable solutions is given by
 \begin{eqnarray}
&&1 = \frac{\lambda_{\rm crit\,1}}{|\lambda |}\frac{1}{1+\frac{Q_1^2}{4}\left( 
\frac{\lambda_{\rm crit\,1}}{\lambda}\right)^2} + \frac{\lambda_{\rm crit\,2}}
{|\lambda |}\frac{1}{1+\frac{Q_2^2}{4}\left(\frac{\lambda_{\rm crit\,1}}
{\lambda}\right)^2}\nonumber \\ && = \frac{2}{2}\frac{\lambda_{\rm crit}}
{|\lambda |}\frac{1}{1+\frac{(2Q)^2}{4}\left( 
\frac{\lambda_{\rm crit}}{2\lambda}\right)^2}= \frac{\lambda_{\rm crit}}
{|\lambda |}\frac{1}{1+\frac{Q^2}{4}\left(\frac{\lambda_{\rm crit}}{\lambda}
\right)^2}\,.
\end{eqnarray}
Thus, if for example $Q=1$, the two components with $Q=2$ would be deemed rather
stable, whereas the compound disk is on the brink of instability. This means
in particular for a galactic disk with a star and gas disk that only
\begin{equation}
Q_* > 1 \qquad {\rm and} \qquad Q_{\rm g} > 1
\end{equation}
ensures dynamical stability. The reverse is also true: if one of the subsystems
is unstable, the entire disk is dynamically unstable. This is in my view a
worrying aspect of the concept of Kennicutt (1989) of dynamically unstable gas
disks.  

\section{The prototype Sc galaxy NGC\,6946}

NGC\,6946 has been studied in great detail. The radial distributions of atomic
and molecular hydrogen, the radial optical surface brightness profile and
estimates of the radial variation of the star formation rate as deduced from
H$_\alpha$ observations can be found in Tacconi \& Young (1986). In the
following I have converted densities of molecular hydrogen to the CO--to--H$_2$
conversion factor of Dame (1993). The rotation curve of NGC\,6946 has been
observed by Carignan et al.~(1990), who also provide a mass model comprising a
disk and a dark halo component. All the data are summarized in Table 1. The gas
densities have multiplied by a factor of 1.4 in order to take account of the
heavy elements. In order to determine the $Q$ parameters estimates of the
velocity dispersions are required. For the interstellar gas I assume following
Kennicutt (1989) the ubiquitously found value of 6 km/s (Dickey et al.~1990).
Kamphuis \& Sancisi (1993) derive from their HI observations of NGC\,6946 a
velocity dispersion of the HI gas of about 13 km/s in the regions of the optical
disk (R$_{25}$ = 16 kpc), which drops to 6 km/s outside the optical disk.
Ferguson et al.~(1998) have pointed out that, if the higher value is used to
determine the stability parameter, $Q_{\rm g}$ does not drop below the threshold
of instability. However, within about half the optical radius the interstellar
gas is dominated by molecular hydrogen and the velocity dispersion of molecular
clouds is typically 5 km/s (Gammie, Ostriker, \& Jog 1991). I have thus chosen
the velocity dispersions given in Table 1. The radial stellar velocity
dispersions are not known, but can be estimated from the vertical hydrostatic
equilibrium condition of the disk and assumptions about the form of the velocity
ellipsoid as explained in Fuchs \& von Linden (1998). The resulting stability
parameters including corrections for the finite thickness of the disk are shown 
in Table 1.
\begin{table}
\caption{Radial variation of stability parameters \\ of the disk of NGC\,6946 
(d=10 Mpc).}
\begin{tabular}{lrccccccc}
\tableline
R & $\Sigma_*$ & $\kappa$ & $\sigma_{\rm U\,*}$ & $Q_*$ & $\Sigma_{\rm g}$ &
$\sigma_{\rm g}$ & $Q_{\rm g}$ & $Q_{\rm g,min}$ \\
\tableline
5  & 133 & 50 & 114 & 2.5 & 38 & 6 & 0.6 & 0.9 \\
10  & 55 & 23 & 89 & 2.2 & 21 & 6 & 0.5 & 1.1 \\
12  & 38 & 20 & 76 & 2.3 & 17 & 6 & 0.5 & 1.1 \\
14  & 27 & 17 & 66 & 2.4 & 14 & 6 & 0.5 & 1.1 \\
16  & 19 & 15 & 52 & 2.3 & 6 & 6 & 1.0 & 1.0 \\
18  & 13 & 13 & 44 & 2.6 & 5 & 6 & 1.1 & 1.0 \\
20  & 9 & 12 & 39 & 3.0 & 5 & 6 & 1.0 & 1.0 \\
kpc & $\frac{{\rm M}_\odot}{{\rm pc}^2}$ & $\frac{{\rm km\,s^{-1}}}{{\rm kpc}}$
 & ${\rm km\,s^{-1}}$ &  & $\frac{{\rm M}_\odot}{{\rm pc}^2}$ & ${\rm km\,
 s^{-1}}$ & & \\
\tableline
\tableline
\end{tabular}
\end{table}
As can be seen from Table 1 there is a distinct drop of the stability parameter
of the gas disk below 1 at R = 16 kpc, which coincides with the outer boundary
of the HII region disk. I conclude from this discussion like Kennicutt (1989)
that the massive star forming disk of NGC\,6946 is indeed dynamically unstable.

\section{Numerical Simulations}

In order to follow the onset of instability into the non--linear regime I have
run together with S.~von Linden numerical simulations of the evolution of an
unstable gas disk embedded in a star disk. The code, originally developed by
F.~Combes and collaborators, implements a two--dimensional star disk (N$_*$ =
38\,000) in which interstellar gas clouds are embedded (N$_{\rm c} \leq$
38\,000). The composite disk is surrounded by rigid bulge and dark halo
potentials. The gravitational potential of the star disk is calculated by a
standard particle--mesh scheme and the inelastic encounters of the gas clouds is
simulated by an elaborate cloud--in--cell scheme, which maintains by coalescence
and fragmentation a steady mass spectrum of the clouds. Details of the
simulations are described in Fuchs \& von Linden (1998). The composite disk was
set up initially axisymmetrically, and the $Q$ parameters
of the stars and the gas were initially $Q_* \approx 2$ and $Q_{\rm g} \approx
 0.5$, respectively. Thus the composite disk was set up dynamically unstable and
 resembles in this the inner parts of the Sc galaxies from Kennicutt's sample.
\begin{figure}[h]  
  \caption{
  cf.~Fig.~7 of Fuchs \& von Linden (1998)\\
  Dynamical evolution of the star and gas disks. On the left--hand side
  of each panel 19\,000 out of 38\,000 stars and on the right--hand side of each
  panel 19\,000 out of 38\,000 interstellar gas clouds are plotted at
  consecutive time intervals. The time is indicated on the left in units
  of 10$^7$ yrs. The spatial size is indicated at the bottom by a bar of 10 kpc
  length.}
  \label{fig2}
\end{figure}
The next steps of the evolution of the disk are shown in Fig.~2. As expected
ring--like density perturbations
appear immediately in the gas disk. The wave lengths of the perturbations are of
the order of the critical wave length of the gas disk, about 2 kpc at R = 10
kpc. The rings fragment into lumps with masses in the range of 10$^4$ to 10$^7$
M$_\odot$. These agglomerates are so heavy that the star disk responds to them
by induced, `swing amplified' spiral structures (Toomre 1981). The potential
troughs of the star disk, on the other hand, begin to trap much of the
interstellar gas, and during their further evolution both the star and gas disks
undergo rather synchronous, repetitive cycles of swing amplified spiral
perturbations. After 5$\cdot$10$^8$ yrs the star disk gets heated up dynamically
by the spiral activity so much that hardly any non--axisymmetric structure is
any longer possible in the disk. In the gas disk, however, there is still a lot
of spiral activity. Since the star disk has become dynamically inactive, the
critical wave lengths of the spiral structures are much smaller, which leads to
a flocculent appearance of the disk.

\section{Discussion}

The numerical simulations show that the disks of Sc galaxies like NGC\,6946 are
in a highly peculiar dynamical state. The reaction of the star disks to
dynamically unstable gas disks is so fierce that they become dynamically hot
within less than a Gyr. On the other hand, Toomre (1990) has argued that Sc
galaxies must have star disks, which are dynamically active, because otherwise 
their morphological appearance would be quite different from what is 
observed. This can be clearly seen in Fig.~2, when one compares the frames
corresponding to, say, 1.4$\cdot$10$^8$ yrs and 4.8$\cdot$10$^8$ yrs with an
optical image of the galaxy which is reproduced in Fig.~3.
\begin{figure} [h]
\epsfig{file=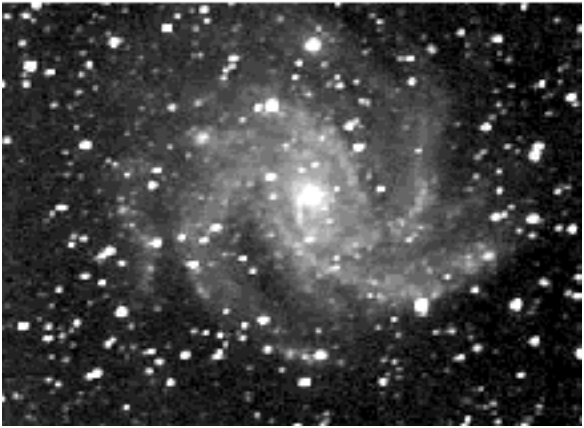}
\caption{Optical image of NGC\,6946 (AURA/NOAO).}
\label{fig3}
\end{figure}
Thus the star disks in Sc galaxies must be effectively cooled dynamically by
star formation of stars on low velocity dispersion orbits. The star formation
rate required to cool the star disks has been estimated by Fuchs \& von Linden
(1998).
They conclude from their numerical simulations that about 40\% of the mass of 
the star disk is required per Gyr in the form of newly born stars to keep the
disk in a steady dynamical state. Interestingly the presently observed star 
formation rate in NGC\,6946 as deduced from the extinction--corrected
H$_\alpha$ surface emissivity (Devereux \& Young 1993) is actually as high as 
the required gas consumption rate. This is in my view not a coincidence, but
has to be explained by theories of the physics of star formation.
On the other hand, this mode of star formation cannot be sustained over 
extended periods, because a galaxy like NGC\,6946 is consuming at the present 
rate nearly its entire gas disk within a Gyr. It will switch presumably soon 
to a more quiescent mode of star formation like in M\,33, NGC\,2403, or
NGC\,7331, where the threshold of dynamical instability is not reached. This is
quite different in young galactic disks, where the star disks are less massive 
and there is an ample reservoir of interstellar gas (Fuchs et al.~2001, in
preparation).
 
\acknowledgements
I am grateful to J.~Gallagher for helpful discussions.


\begin{references}

\reference Biermann, P.L. 1975, in IAU Symp. No. 69, Dynamics of Stellar
Systems, ed. M. Hayli (Dordrecht: Reidel), 321

\reference Binney, J., Tremaine, S. 1987, Galactic Dynamics (Princeton:
Princeton Univ. Press)

\reference Carignan, C., Charbonneau, P., Boulanger, F. et al. 1990 \aap, 234,
43

\reference Dame, T.M. 1993, in AIP Conf. Proc. 278, Back to the Galaxy,
ed. S.S. Holt \& F. Verter (New York: AIP), 267

\reference Devereux, N.A., Young, J.S. 1993, \aj, 106, 948

\reference Dickey, J.M., Murray Hanson, M., Helou, G. 1990, \apj, 352, 522

\reference Elmegreen, B.G. 1995, \mnras, 275, 944  

\reference Ferguson, A.M.N., Wyse, R.F.G., Gallagher, J.S. et al. 1996, in The
Interplay Between Massive Star Formation, the ISM and Galaxy Evolution, ed. D.
Kunth, B. Guideroni, M. Heydar-Malayeri, \& Trinh Xu Thuan (Gif-sur-Yvette:
Edition Frontieres), 557 

\reference Ferguson, A.M.N., Wyse, R.F.G., Gallagher, J.S. et al. 1998, \apj,
506, L19

\reference Fuchs, B., von Linden, S. 1998 \mnras, 294, 513

\reference Gammie, C.F., Ostriker, J.P., Jog, C. 1991, \apj, 378, 565 

\reference Jog, C., Solomon, P.M. 1984 \apj, 276, 114

\reference Kamphuis, J., Sancisi, R. 1993, \aap, 273, L31 

\reference Kennicutt, R.C. 1989, \apj, 344, 685

\reference Romeo, A.B. 1992, \mnras,  256, 307

\reference Safronov, V.S. 1960, Ann. d'Astrophys., 23, 979

\reference Tacconi, L.J., Young, J.S. 1986, \apj, 308, 600
  
\reference Toomre, A. 1964, \apj, 139, 1217

\reference Toomre, A. 1981, in The Structure and Evolution of Normal Galaxies,
ed. S.M. Fall \& D. Lynden-Bell (Cambridge: Cambridge Univ. Press), 111

\reference Toomre, A. 1990, in Dynamics and Interactions of Galaxies, ed. R.
Wielen (Berlin: Springer), 292

\end{references}
\end{document}